# Design & Development of the Graphical User Interface for Sindhi Language

IMDAD ALI ISMAILI*, ZEESHAN BHATTI**, AND AZHAR ALI SHAH**

## ABSTRACT

**This paper describes the design and implementation of a Unicode-based GUISL (Graphical User Interface for Sindhi Language). The idea is to provide a software platform to the people of Sindh as well as Sindhi diasporas living across the globe to make use of computing for basic tasks such as editing, composition, formatting, and printing of documents in Sindhi by using GUISL. The implementation of the GUISL has been done in the Java technology to make the system platform independent. The paper describes several design issues of Sindhi GUI in the context of existing software tools and technologies and explains how mapping and concatenation techniques have been employed to achieve the cursive shape of Sindhi script.**

## 1. INTRODUCTION

Today computers play very important role in the daily life of common people, and it is observed that the use of computers in countries other than developing ones is substantial in every walk of life. It is a fact that the more the developing countries transform their systems to adapt the use of computers the more progress and development they would achieve and hence would reduce the so called gap of 'digital divide'. The public use of computers, however, is some how dependent on the support for the regional languages of the countries.

There are thousands of languages being spoken throughout the world, and it has been noticed that the English speaking countries have an edge over other nations where English is not as their native language in use of computers. There are many countries which have adopted their native language in use of computers and succeeded to cope with computing problems. There are some countries where multiple languages are spoken such as India, Sri Lanka and Pakistan. In Pakistan Urdu, Punjabi, Sindhi, Pashto, Blochi and Siraiki etc. are spoken [1].

Sindhi is the official language of Sindh province of Pakistan and it is also one of the 22 constitutionally recognized languages of India. Sindhi is spoken by an estimated 34.4 million people in Pakistan (3rd most spoken language) and about 2.8 million people in India (majority of them being those who migrated from Sindh to India in the wave of the partition of the subcontinent in 1947). During and after the partition of the subcontinent many Sindhi families (both Hindu and Muslims) also moved abroad for permanent settlement in many countries across

* Pro Vice Chancellor, Mirpur Khas Campus, University of Sindh, Jamshoro,
** Assistant Professor, Institute of Information & Communication Technologies, University of Sindh, Jamshoro.

the globe; thus forming a significant community of Sindhi diasporas [2].

Sindhi language is classified as an Indo-Aryan language belonging to Indo-Iranian branch of the Indo-European language family. Though the original script of Sindhi language found in the remnants of Mohen-jo-Daro (illustrator of some 5000 bc Indus valley civilization) has yet not been interpreted, the preliminary system of writing appeared before 8th century AD (i.e. before the advent of Islam in Sindh in 711 AD). The contributions of Sufi poets like Shah Abdul Latif Bhitai, Sachal Sarmast and Sami further popularized Sindhi language in literary circles in between 14th and 18th century [3]. Since this time, the literature in Sindhi language has grown significantly and currently Sindhi is being used as medium of instruction in Sindhi majority public sector schools of province of Sindh from primary to secondary level (matriculation). Dozens of daily Sindhi newspapers are being published along with several dedicated Sindhi TV channels. Besides this Sindhi is also one of the two languages used by Government of Pakistan to issue computerized national identity cards to its citizens.

Though the proper use of Sindhi language in all of the above mentioned spheres requires its proper adaptation and progress in terms of modern computer based standards and models; however, too little work has yet been done for the standardization of Sindhi computing as well as development of computer based models of Sindhi script, speech and language [4]. The major issues in Sindhi computing as investigated and observed are due to the following reasons:

(i) No localized Sindhi Software available.

(ii) Non-availability of compatible Sindhi fonts for different operating systems with standard Unicode format.

(iii) Complex process of enabling Sindhi through "Regional and Language Settings".

(iv) Non-compatibility of Sindhi keyboard layout for different operating systems.

(v) Lack of computer literature and books in Sindhi.

(vi) No educational programs in Sindhi Computing at School and College levels.

(vii) Research activities at infancy in Sindhi computing.

(viii) Communication and coordination gap among researchers and professionals etc.

As far as we know, so far, neither the provincial government of Sindh nor the federal government of Pakistan has initiated any significant project to fund the research on Sindhi language processing. Not only this but the complete failure of the government to control copy right system and eliminate the use of pirated software has also contributed to the lack of willingness of developers to commit any resources for the research and development of Sindhi language. As a result of this there are only a few legacy tools being used for Sindhi desktop publishing which vary from one publisher/individual to another publisher/individual. Almost all of these tools adapt English-based word processing systems and make use of their specific code pages (mostly based on ASCII scheme) for Sindhi character set. To the best of our knowledge, the development of a Sindhi-based GUI for the processing of Sindhi language has not yet been reported. This paper takes a step towards this aim and presents the design and development of a GUI for SL (Sindhi Language). The rest of the paper is organized as follows: Section 2 presents the background and motivation along with a thorough discussion of the major issues in terms of Sindhi computing and also in terms of the design of the GUISL; Section 3 presents the design and development of the GUISL and finally Section 4 concludes the paper with added description of the future work.

## 2. GUISL: DESIGN ISSUES

Historically, Sindhi language has adopted a variety of writing systems based on the innovations and preferences of particular regional communities. Most of the names of the ancient scripts are either based on the names of the cities/towns (e.g., Khudabadi, Thattai, etc.) or the names of the communities (e.g. Memonki, Luhaniki, Khojiki, Devnagri, Gurmukhi, Perso-Arabic etc.) [8]. During the colonial period, owing to the Sansakrit related nature of Sindhi, some British scholars advocated for the promotion of Devnagri script to be used for Sindhi. To this aim they managed to publish the translation of Bible and Sindhi-English dictionary using Devnagri script in 1849- 1850 [10]. This move was opposed by some government employees who were only familiar with Perso-Arabic script and their side was backed by some British officials including Captain Sir Richard Francis Burton. The matter got referred to the Court of Directors of the British East India Company.

Based on the fact that the Muslim names were not properly written with Devnagri, the use of the Perso-Arabic script was recommended by the court. Following this decision, a team of scholars including Sir Burton, Munshi Thanwardas and Mirza Sadiq Ali Beg worked on the standardization of Perso-Arabic based script for Sindhi language. The standardization of this script was completed in 1852 and it consisted of 52 letters derived from Perso-Arabic scripts with addition of dots and lines to fully represent all of the Sindhi sounds. This script is currently being used in Sindh and abroad and is based on Arabic Nashk style of writing [6,10-11]. Figs. 1-2 show the list of these characters.

Short vowels and some additional vocalic and consonantal features are also represented through diacritical (Zeer, Zaber, Peush, etc.) marks in Sindhi [3,8] as listed in Fig. 3. The diacritics (also known as aerab) are optionally used in writing, however, for our system development they have not been used for the sake of simplicity and readability.

Thus, the design of GUISL involves numerous key factors that need to be addressed for designing the interface. Sindhi is written and read from right-to-left direction, therefore the system should have the ability to display text from right-to-left and everything of GUI (toolbar, status bar, writing direction, menus, tool tips, popup menus etc.) should be presented from right-to- left for accommodating the Sindhi users.

According to the Unicode standard the characters should be inserted and stored in a simple logical sequence [5]. If we wish to write the word "سنڌي" (Sindhi), the keys would be pressed in succession corresponding to the characters making up the word "سنڌي" (Sindhi). The text file would store the Unicode codes of these characters in order as shown below.

| 0633 | 0646 | 068C | 06CC |
|---|---|---|---|
| س | ن | ڌ | ي |
| {Seen} | {Noon} | {Dhal} | {Yeh} |

To display the text in an appropriate order of Sindhi the system needs to correspond to each of the successive code to be displayed in the right-to-left order however it would be displayed into left-to-right as shown below. This is what happens with Roman Script making it straight forward and simple to render [4-5,11].

| | | | | | | | | | |
|---|---|---|---|---|---|---|---|---|---|
| ث | ٺ | ٽ | ٿ | ت | ڀ | پ | ٻ | ب | ا |
| sē | ṭhē | ṭē | thē | tē | bhē | pē | ḅē | bē | alif |
| ڌ | د | خ | ح | ڇ | چ | ڃ | ڄ | جھ | ج |
| dhāl | dāl | khē | hē | chē | cē | ñē | jē | jhim | jim |
| ص | ش | س | ز | ڙ | ر | ذ | ڏ | ڊ | ڍ |
| ṣād | shin | sin | zē | ṛē | rē | zāl | ḍhāl | ḍāl | dāl |
| ک | ڪ | ق | ڦ | ف | غ | ع | ظ | ط | ض |
| khē | kāf | qāf | phē | fē | ghayn | ảyn | zōē | ṭōē | ẓād |
| ھ | و | ڻ | ن | م | ل | ڱ | ڳ | گھ | گ |
| hē | vāō | ṇūn | nūn | mim | lām | ṇē | gē | ghāf | gāf |
| | | | | | | | | ي | ه |
| | | | | | | | | yē | hē |

FIG. 1. STANDARD SINDHI CHARACTERS COMMONLY USED IN SINDHI SCRIPT

| ء | لا | آ | ئ | ؏ | ۾ |
|---|---|---|---|---|---|
| Hamza | LamAlif | AlifMad | HamzaYeh | Aien | Min |

FIG. 2. ADDITION SINDHI CHARACTER USED IN SINDHI LANGUAGE

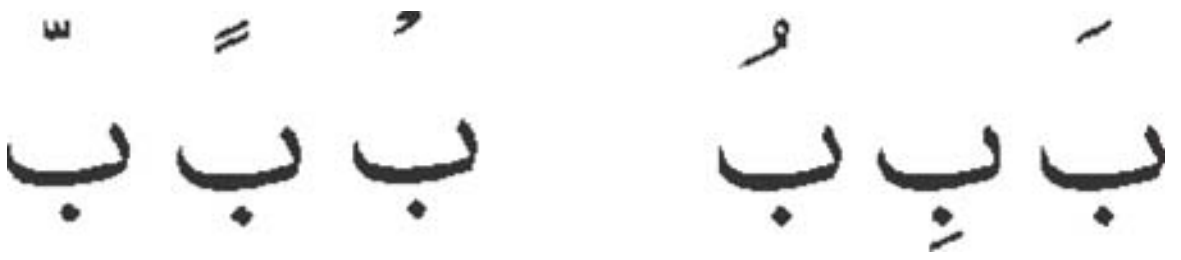

FIG. 3. SINDHI DIACRITICS

ي ذن س

Thus it needs reordering or grouping with one-to-many relationship between characters. As Sindhi language is a cursive like Arabic in its written format [6] the above text is unacceptable as neither the ordering is correct nor the shapes of the characters as per cursive nature of the script.

The system should be able to display Sindhi character properly with its cursive forms while ligature used to form Sindhi words. The development of the GUISL should provide easy access for ordinary user and without any additional requirements such as Sindhi Fonts, Keyboard Layout setup, Regional Language Setting etc. These design issues pose impediments in the development of the Sindhi software system.

## 3. Design & development of the GUISL

Three major issues are analyzed to design and standardize the Sindhi user interface: Unicode of characters, keyboard layout, and Sindhi cursive form. Unfortunately, there is no ordering (sorting) of Sindhi Unicode characters in the Unicode plane and all the characters are scattered and mixed with Arabic and Farsi character set as shown in Fig. 4 (Sindhi characters are circled highlighting their positions in the plane). The absence of a sequence and segregated set of character coding leads to develop ad-hoc based systems for process of Sindhi computing. The first step involves changing the orientation of all the GUI elements (toolbar, status bar, writing direction, menus, tool tips, popup menus etc.) in system from right-to-left allowing the initial, middle and last representations of the characters to be rendered appropriately with proper cursive for the ligature to form the word [9].

### 3.1 Unicode Based Sindhi Characters Coding

A static class was created in Java language, declaring all the Unicode based Sindhi characters to be used in GUISL as shown in pseudo-code given in the Algorithm-1. The list of characters along with their variable names mapped with Unicode of each character is shown in Table 1.

### 3.2 Sindhi Cursive Form

Sindhi script has a cursive form similar to that in Arabic, that is, the letters in the Sindhi script join together into units to form words. Sindhi script also has context sensitive glyph shaping; depending on whether the character joins a word in the initial, medial or final position or is isolated taking a different shape as shown in Table 2. Nonetheless, cursiveness, ligation and context sensitivity are rendering related issues and the output shapes of characters may vary with context, their internal encoding remains unchanged. For example, the letter È{b:Beh} may take multiple shapes but its internal encoding is always U+0628 as shown in Table 2. Therefore, these properties have no implication on collation [5].

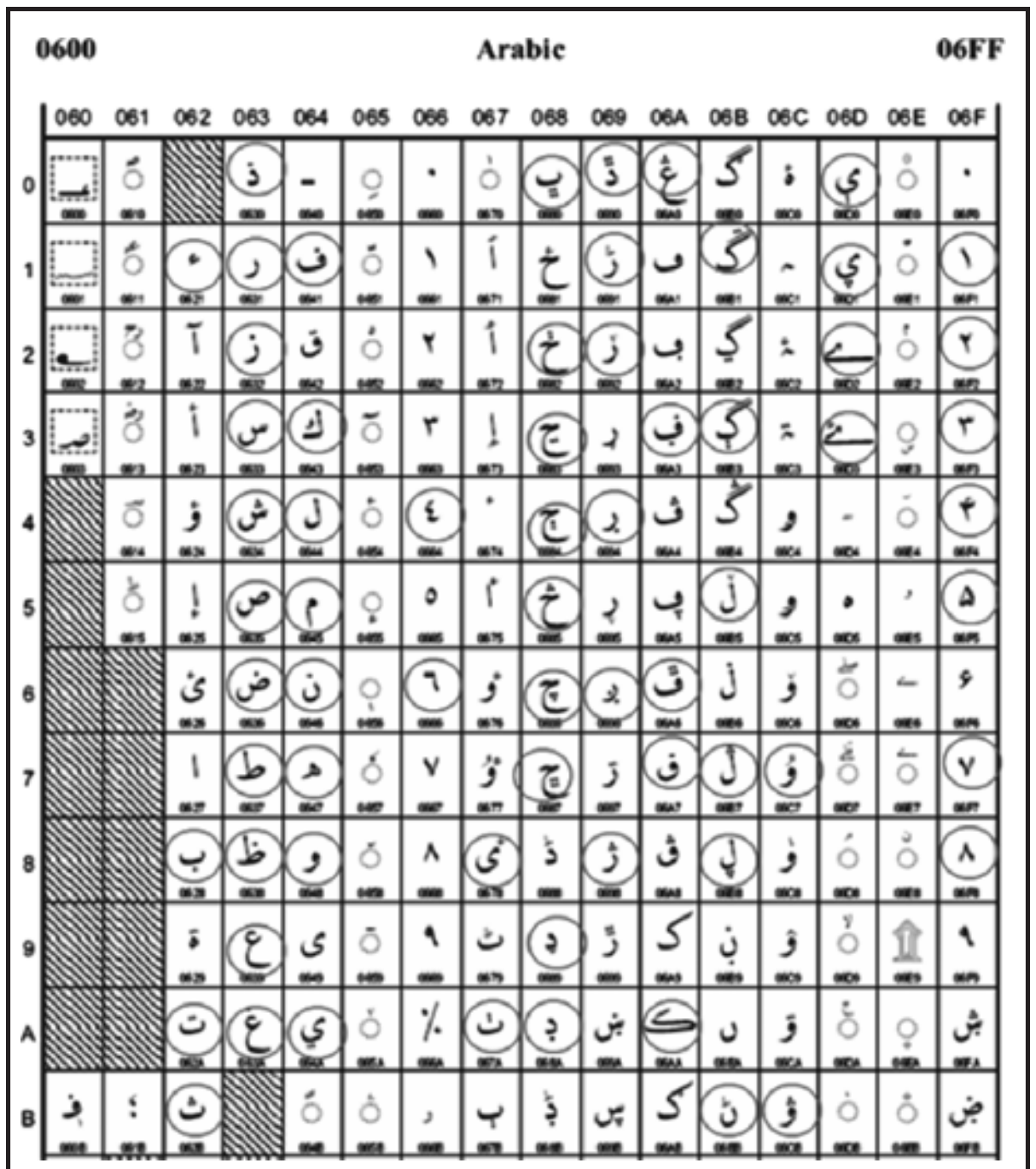

*FIG. 4. ARABIC/SINDHI UNICODE PLANE: THE ANNOTATION OF CIRCLES INDICATES THE RELATIVE POSITION OF SINDHI ALPHABET CHARACTERS IN THE PLANE (FOR CLARITY OF UNICDE VALUES PLEASE REFER TO TABLE 1). (SOURCE: HTTP://WWW.UNICODE.ORG/ CHARTS/)*

**TABLE 1. LIST OF SINDHI CHARACTER, ITS CORRESPONDING UNICODE AND ITS VARIABLE NAME USED IN THE SYSTEM**

| Variable Name | Unicode | Sindhi Letter | Veriable Name | Unicode | Sindhi Letter |
|---|---|---|---|---|---|
| sheen | 0634 | ش | alifMadA | 0622 | آ |
| swad | 0635 | ص | alif | 0627 | ا |
| dad | 0636 | ض | beh | 0628 | ب |
| toye | 0637 | ط | beeh | 067B | ٻ |
| zoye | 0638 | ظ | peh | 067E | پ |
| aieen | 0639 | ع | beheh | 0680 | ڀ |
| ghain | 063A | غ | the | 062A | ت |
| feh | 0641 | ف | theh | 067F | ٿ |
| peheh | 06A6 | ڦ | mytheey | 067D | ٽ |
| qaf | 0642 | ق | tteheh | 067A | ٺ |
| Kaf | 06AA | ڪ | ttay | 062B | ث |
| keheh | 06A9 | ک | jeem | 062C | ج |
| gaf | 06AF | گ | dyeh | 0684 | ڄ |
| geuh | 06B3 | ڳ | nyeh | 0683 | ڃ |
| ngoeh | 06B1 | ڱ | cheh | 0686 | چ |
| lam | 0644 | ل | cheheh | 0687 | ڇ |
| meem | 0645 | م | hah | 062D | ح |
| noon | 0646 | ن | khay | 062E | خ |
| rnoon | 06BB | ڻ | dal | 062F | د |
| waw | 0648 | و | dahal | 068C | ڌ |
| heh | 06BE | ھ | dhal | 068F | ڏ |
| hamza | 0621 | ء | ddal | 068A | ڊ |
| Yeh | 064A | ي | ddahal | 068D | ڍ |
| yehSmall | 06C1 | ہ | zal | 0630 | ذ |
| yehHamza | 0626 | ئ | reh | 0631 | ر |
| Min | 06FE | ۾ | rdeh | 0699 | ڙ |
| sindhi Ampersand | 06FD | ۽ | zeh | 0632 | ز |
| | | | sceen | 0633 | س |

Algorithm 1: Pseudo code of the static class used for the declaration of Sindhi characters and initialized with their corresponding Unicode characters.

```
<package> <guisl.com.snd>
<public><static><class><class.name>
<public><static><data type: char><variable.name:alif> = <value: '\u0627'>
<public><static><data type: char><variable.name:beh> = <value: '\u0628'>
<public><static><data type: char><variable.name:beeh> = <value: '\u067B'>
<public><static><data type: char><variable.name:peh> = <value: '\u067E'>
<public><static><data type: char><variable.name:beheh> = <value: '\u0680'>
<public><static><data type: char><variable.name:the> = <value: '\u062A'>
<public><static><data type: char><variable.name:theh> = <value: '\u067F'>
<public><static><data type: char><variable.name:ttay> = <value: '\u062B'>
. . .
. . .
. . .
. . .
<end of class>
```

**TABLE 2. MULTIPLE GLYPH FORM OF SINDHI CHARACTERS**

| Unicode | Sindhi Name Form | Isolated Glyph Form | Last Glyph Form | Medial Glyph Form | Initial Glyph Form |
|---|---|---|---|---|---|
| U+0628 | BEG | ب | ـب | ـبـ | بـ |
| U+0646 | NOON | ن | ـن | ـنـ | نـ |
| U+064A | YEH | ي | ـي | ـيـ | يـ |
| U+062D | HAH | ح | ـح | ـحـ | حـ |
| U+0633 | SEEN | س | ـس | ـسـ | سـ |
| U+0635 | SAD | ص | ـص | ـصـ | صـ |
| U+0637 | TAH | ط | ـط | ـطـ | طـ |

The Pseudo code shown in Algorithms 2-3 illustrates the main steps performed on each GUI object for displaying Sindhi text from example 1 and 2 respectively. Each Sindhi character is mapped to a corresponding static variable (Table 1) which is then concatenated with each other to form a Sindhi word as illustrated in examples 1 and 2 below:

Sindhi cursive form: سنڌي {Sindhi}

Indiviual Characters س ن ڌ ي {s: Seen} {n : Noon} {dh: Dhal} {y: Yeh}

Example 1: Sindhi Word to be formed from the variable joining

Algorithm 2. Pseudo code of the function used for the mapping of the Sindhi character from example 1 with its corresponding Unicode (from Table 1), concatenated with each other to form a Sindhi ligature.

```
<function> <function.name>
<variable>= String. concatenate(
""+<variable: sceen> + <variable:noon> +
<variable:dahal> + <variable:yeh>)
<GUI Object> {Change Orientation to
RIGHT.TO.LEFT}
<GUI Object>.setText(<variable>)
<end of function>
```

Sindhi cursive form:

يونيڪوڊ تحت سنڌي لفظن جو خطاط{Unicode Based Sindhi Word Processor}

Example 2: Sindhi Word to be formed from the variable joining

Algorithm 3. Pseudo code of the function used for the mapping of the Sindhi string from example 2 with its corresponding Unicode (from table 1), concatenated with each other to form a Sindhi statement.

```
<function> <function.name>
<variable 1>= String. concatenate(""+<variable:yeh>+
<variable:waw>+<variable:noon>
+<variable:yeh>+<variable:kaf>+<variable:waw>+
<variable:ddal>)
<variable 2>= String. concatenate(""+ <variable:the>+
<variable:hah> + <variable:the>)
<variable 3>= String. concatenate(""
+<variable:sceen>+<variable:noon>+<variable:dahal>+
<variable:yeh>)
<variable4>= String. concatenate( ""+
<variable:lam>+<variable:feh>
+<variable:zoye>+<variable:noon>)

<variable 5>= String. concatenate(""+ <variable:jeem>+
<variable:waw>+" " + <variable:khay>
+<variable:toye>+<variable:alif>+<variable:toye>)
<variable 6>= String. concatenate( ""+<variable 1> + <variable
2> + <variable 3> +
+ <variable 4> +" "+ <variable 5>)
<GUI Object> {Change Orientation to RIGHT.TO.LEFT}

<GUI Object>.setText(<variable 6>)
<end of function>
```

The Algorithm-3 illustrates the Pseudo.code to create a Sindhi String with multiple words being concatenated to form a statement. First each word is concatenated separately and assigned to variable and then all the variables are concatenated to create a complete statement.

## 3.3 GUI in Sindhi Language

Using the variable joining method described in previous section Sindhi GUI architecture was designed as shown in Fig. 5. The system architecture contains the basic GUI elements grouped under the relevant menu. The main menu groups are written in both English and Sindhi script for ease of understanding.

Fig. 6(a-e) demonstrate the various user interface menus with Sindhi script developed for the GUISL. The Menus have been designed with Tab-Based layout instead of conventional Drop-Down menus. The File Menu shown in Fig. 6(a) contains the basic tools regarding the input output operations such as نئون فائيل (New File), فائيل کوليو (Open File), محفوظ ڪريو(Save File), etc.

The Fig. 6(b) shows the درستگي (Edit) menu which contains the tools regarding the basic editing options such as واپس ڪريو (Undo), ٻيهر ڪريو (Redo), ڳولها (Find), etc. Similarly the لغت (Dictionary) menu shown in Fig. 6(c) contains the options for loading various dictionary types such as Sindhi to English Dictionary, English to Sindhi Dictionary, Computer, Medical and Business Sindhi Dictionaries.

Fig. 6(d) contains the on-screen Sindhi keyboard for visual typing using mouse. Two keyboard types have been used in the GUISL system, one with sequential Sindhi keys as shown in Fig. 6(d) and the other is the on-screen Sindhi keyboard layout as shown in Fig. 6(e).

## 3.5 User Convenience

In the GUISL the tooltips have also been written in Sindhi-English format so that the user may get the information on a particular tool both in Sindhi and in English. Fig. 7(a-c) shows the various tooltips in Sindhi-English format. The bi-directional text right-to-left for Sindhi and left-to- right for English is used. The same method of concatenating Unicode variables from Table 1 to English words have been used to display the text. Fig. 7(a) shows the tooltip for the 'Save File'. Similarly Fig. 7(b) displays the tooltip regarding the different dictionary types.

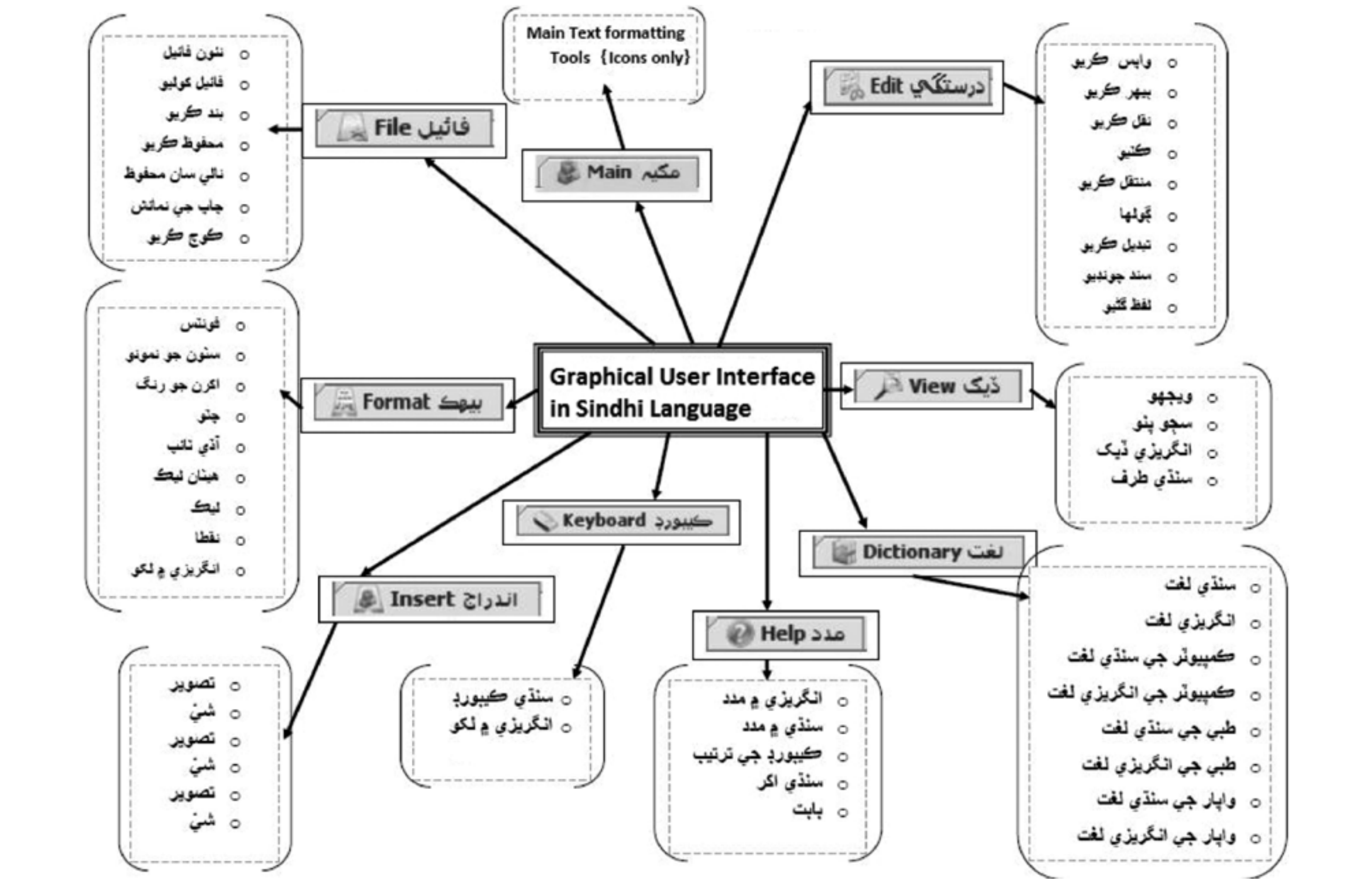

*FIG. 5. THE MAIN ARCHITECTURE OF THE GUI IN SINDHI*

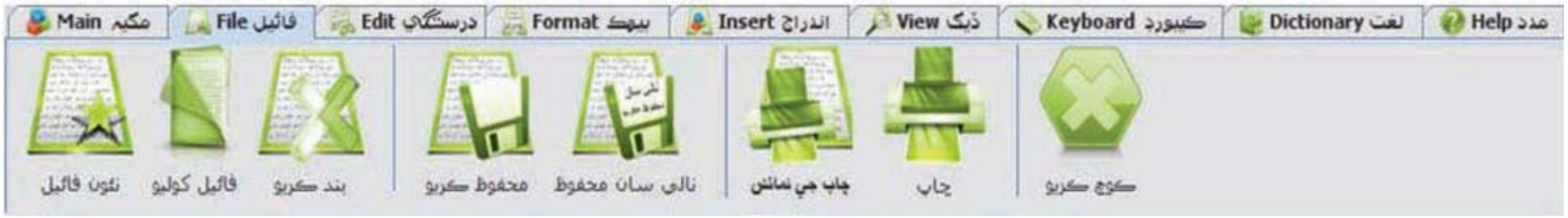

*FIG. 6(A). FILE MENU*

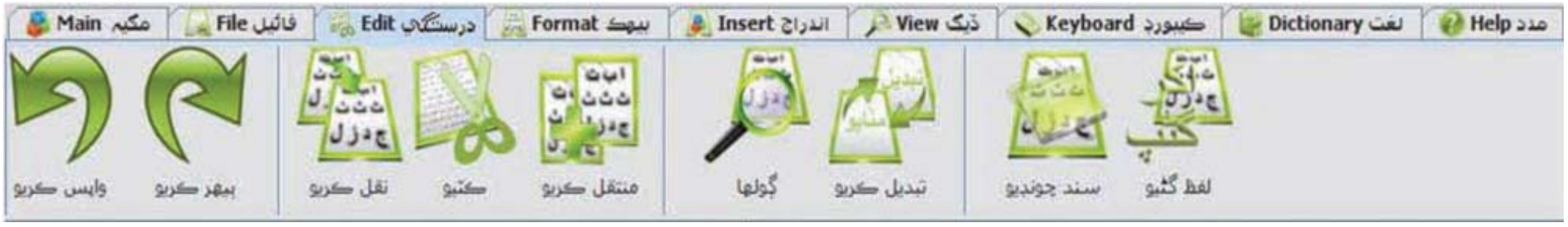

*FIG. 6(B). EDIT MEMU*

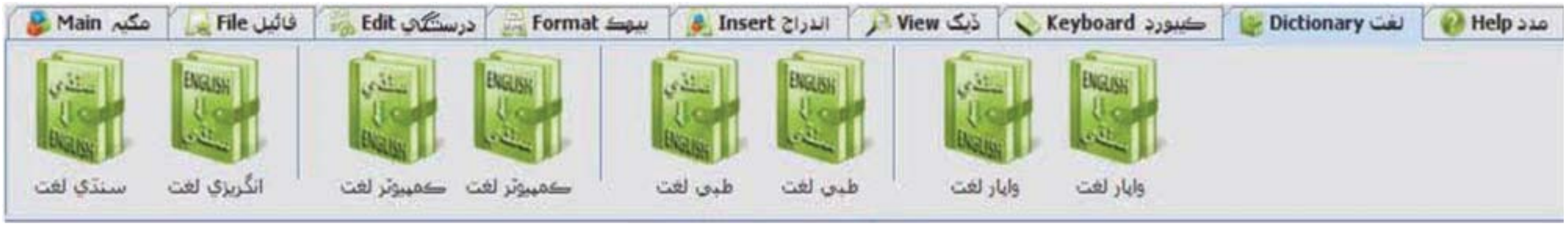

*FIG. 6(C). DICTIONARY MENU*

The Fig. 7(c) displays the tooltip for the 'Undo', 'Redo' and 'Exit' options.

### 3.6 Message and Tool Windows

All the pop-up message and tool windows in the system contain the text in both Sindhi and English language as shown in Fig. 8(a-b). Fig. 8(a) shows the message window that appears before closing the file asking to save the changes both in Sindhi and English.

Fig. 8(b) displays the 'Find and Replace' tool Window have Sindhi-English text in the title bar of the window and in the tabs of Find and Replace options.

Fig. 8(c) shown below is for the task pane that displays the various tools and tasks in the system with Sindhi text on its labels and buttons.

## 4. CONCLUSIONS & FUTURE DIRECTIONS

This paper has concentrated on the issue of designing and development of a GUISL. Various characteristics and issues affecting the use of Sindhi in computing have been discussed such as changing the orientation of the text from right-to-left, ordering/sorting of Sindhi Unicode characters and proper cursiveness for the ligation of words. After successful implementation of the algorithms discussed, the GUI of any application can be defined in Sindhi language. The issues and methods discussed can also be applied on other languages that are written in Perso-Arabic script.

This work is targeted at the provision of a generic programming framework for the development of Sindhi language based applications. Based on the present work we are now in the process of the development of a platform independent Sindhi word processor that includes localised (in Sindhi) supporting tools such as dictionaries, translation/transliteration, spell-checker, calendar and calculators.

## ACKNOWLEDGEMENTS

Authors are indebted to acknowledge the valuable suggestions of the anonymous reviewers which helped in further improving the readability of this article. The constructive discussion with officials of SLA (Sindhi Language Authority), regarding the use of proper technical Sindhi words is also highly appreciated and acknowledged.

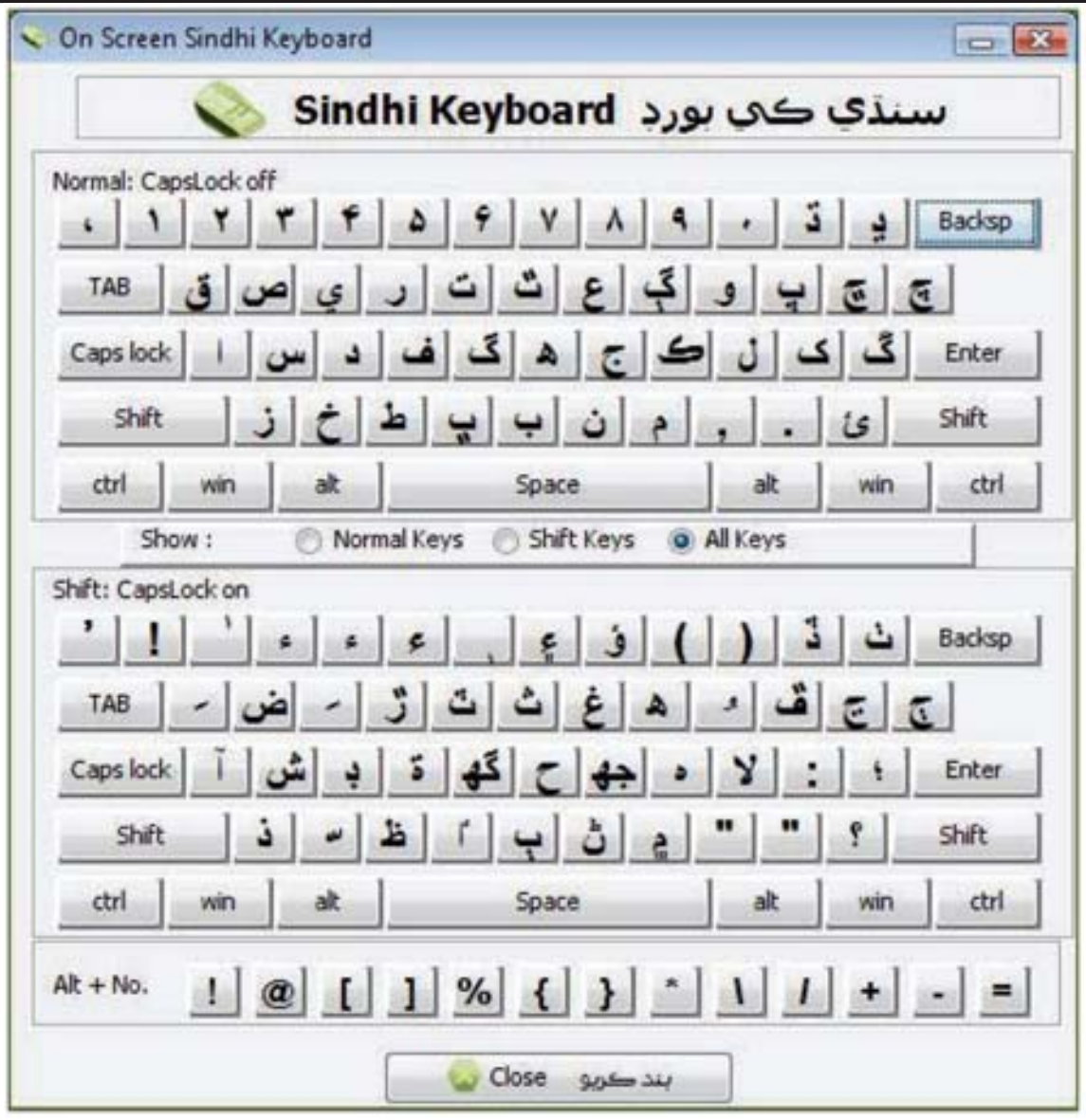

FIG. 6(E). ON.SCREEN SINDHI KEYBOARD

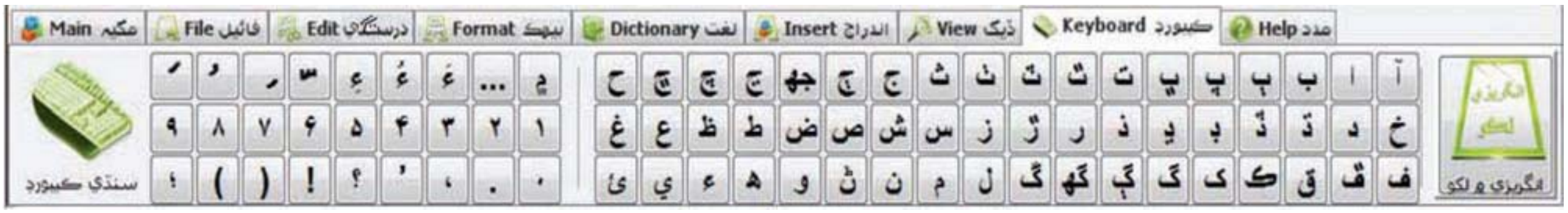

FIG. 6(D). ON-SCREEN SINDHI SEQUENTIAL KEYBOARD

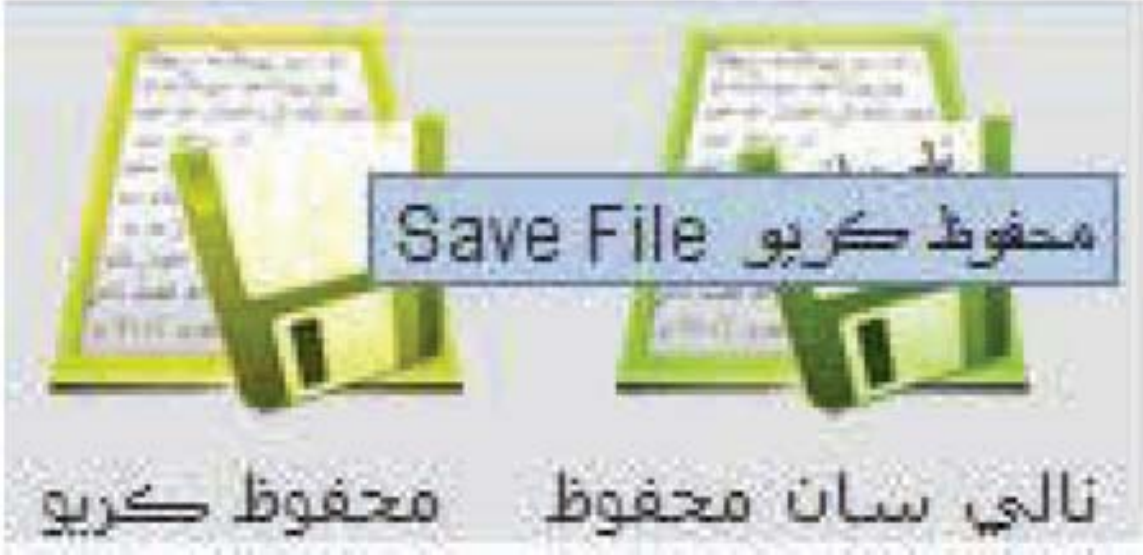

FIG. 7(A). TOOL TIP ON SAVE FILE OPTION

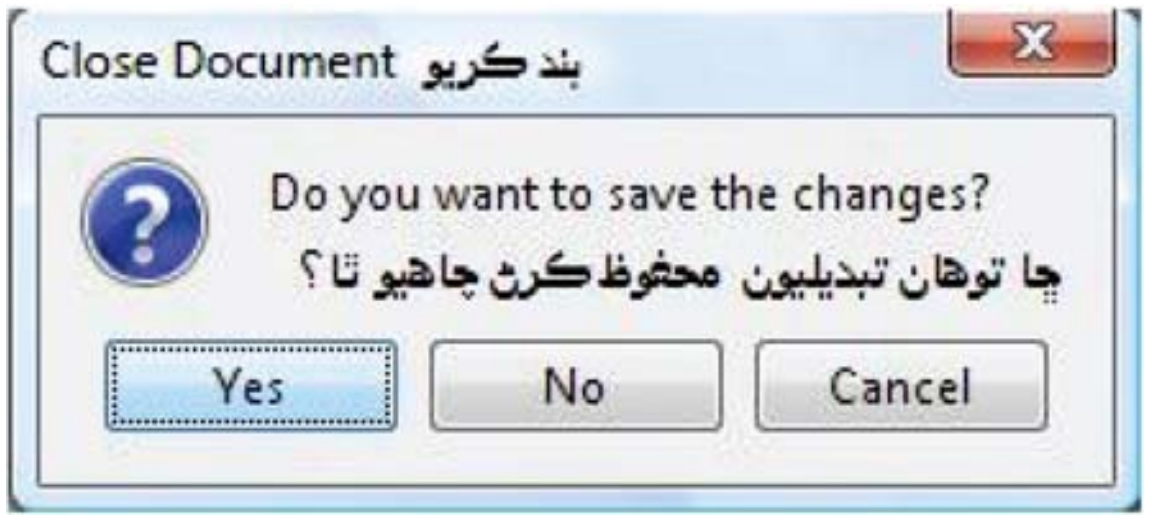

FIG. 8(A). SAVE FILE MESSAGE DIALOG

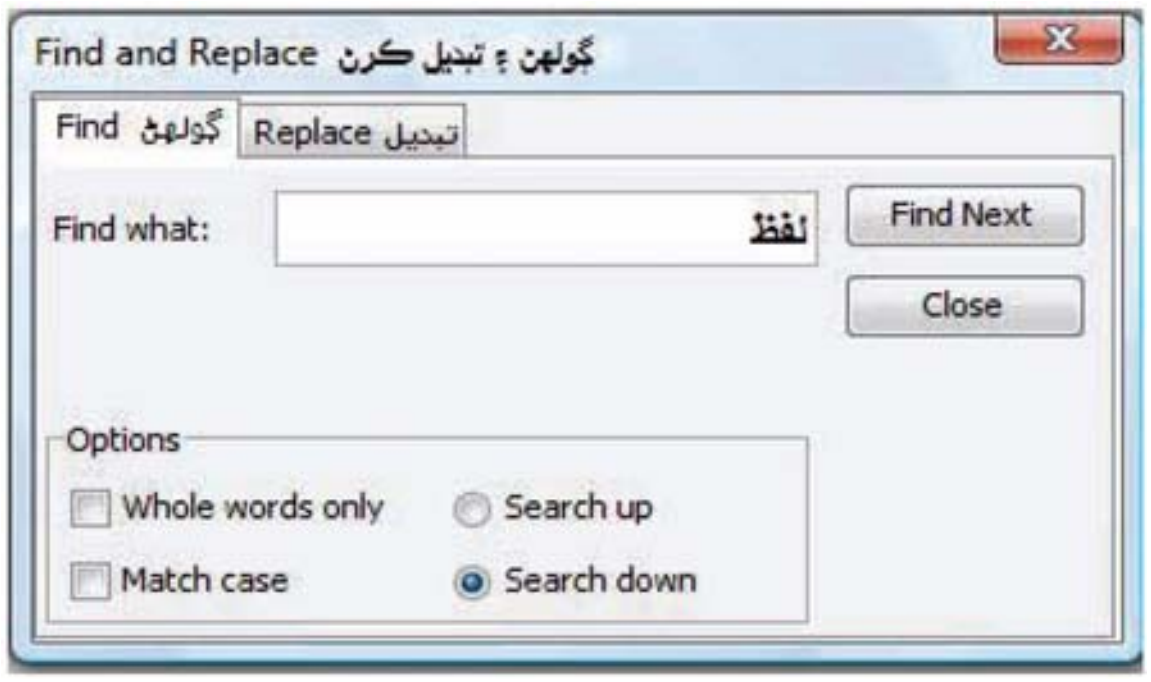

FIG. 8(B). FIND DIALOG

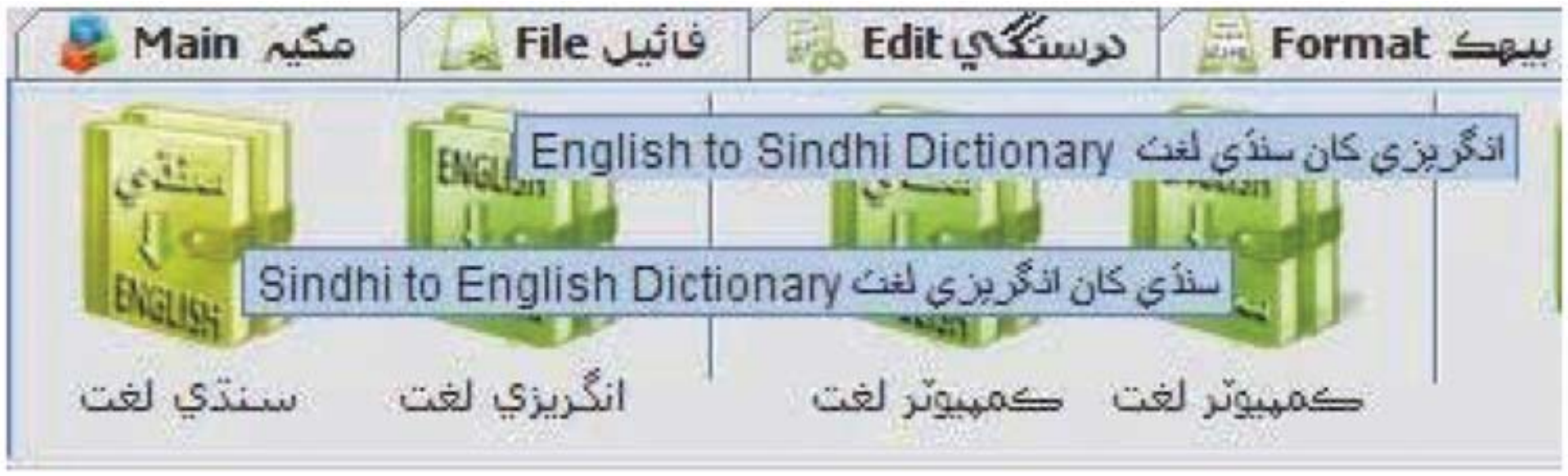

FIG. 7(B). TOOL TIPS ON DICTIONARIES

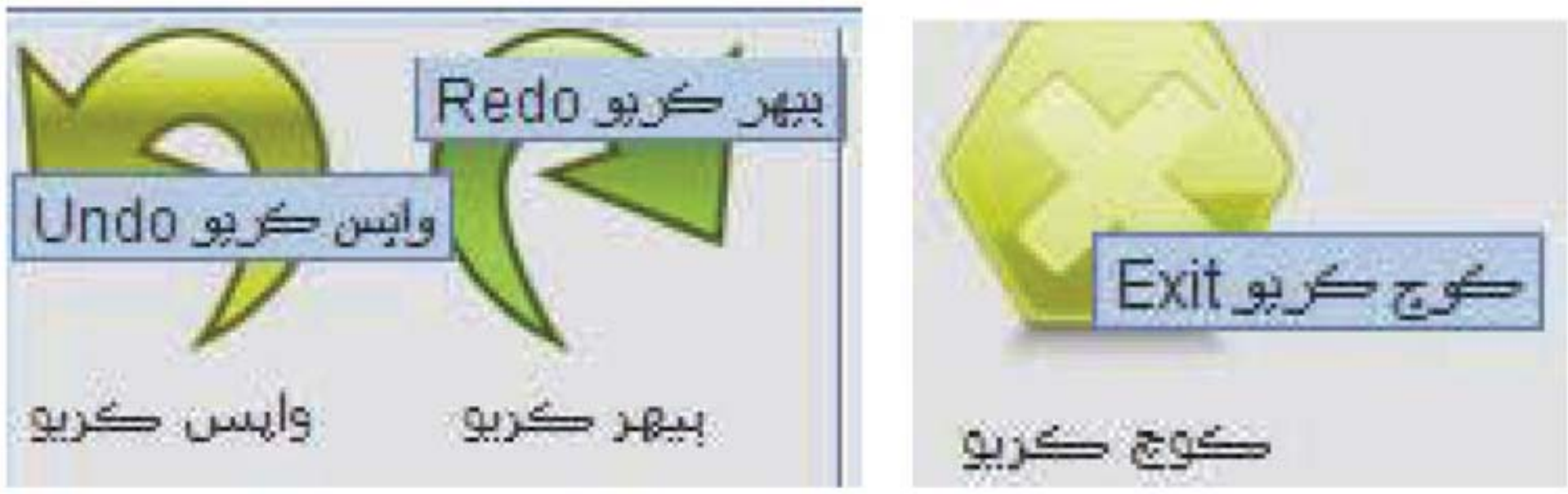

FIG. 7(C). TOOL TIP ON UNDO AND REDO AND EXIT OPTIONS.

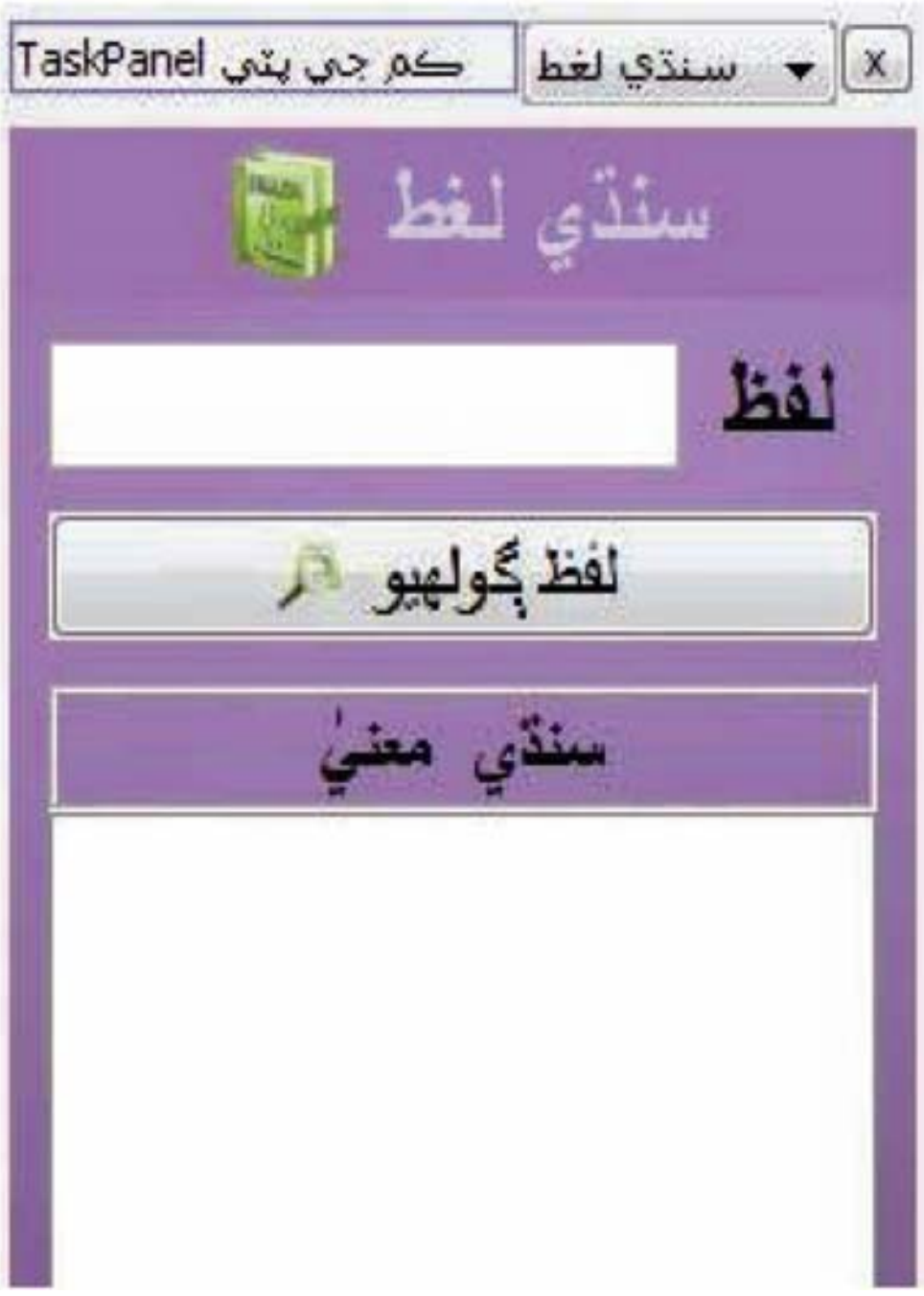

*FIG. 8(C). TASK PANE FOR DICTIONARY*